*electronics*

*Article*

# Estimation during Design Phases of Suitable SRAM Cells for PUF Applications Using Separatrix and Mismatch Metrics

Abdel Alheyasat [1], Gabriel Torrens [1,2], Sebastià A. Bota [1,2] and Bartomeu Alorda [1,3,*]

1 Industrial and Construction Engineering Department, University of the Balearic Islands, 07120 Palma, Spain; heyasat.abdel@uib.es (A.A.); gabriel.torrens@uib.es (G.T.); sebastia.bota@uib.es (S.A.B.)
2 Biosensors, Medical Instrumentation and Data Analysis Group, 07120 Palma, Spain
3 eHealth and Multidisciplinary Telemedicine Research Group, 07120 Palma, Spain
* Correspondence: tomeu.alorda@uib.es

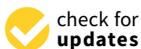

**Abstract:** Physically unclonable functions (PUFs) are used as low-cost cryptographic primitives in device authentication and secret key creation. SRAM-PUFs are well-known as entropy sources; nevertheless, due of non-deterministic noise environment during the power-up process, they are subject to low challenge-response repeatability. The dependability of SRAM-PUFs is usually accomplished by combining complex error correcting codes (ECCs) with fuzzy extractor structures resulting in an increase in power consumption, area, cost, and design complexity. In this study, we established effective metrics on the basis of the separatrix concept and cell mismatch to estimate the percentage of cells that, due to the effect of variability, will tend to the same initial state during power-up. The effects of noise and temperature in cell start-up processes were used to validate the proposed metrics. The presented metrics may be applied at the SRAM-PUF design phases to investigate the impact of different design parameters on the percentage of reliable cells for PUF applications.

## 1. Introduction

IoT devices should provide safe and reliable operation; however, most of the common strategies used to achieve these goals are not suitable when energy consumption and scalability issues are considered. Physical unclonable functions (PUFs) can be used as sources of randomness for the cryptographic modules. Various PUF structures are proposed in the literature: ring oscillator PUFs [1] and SRAM-PUFs [2] are among the most used ones. In addition, there are other interesting alternatives such as DRAM PUFs, chaos-based PUFs, or arbiter PUFs. A more detailed discussion of the possible advantages and disadvantages of such alternatives can be found in references [3–6]. Since most digital circuits contain at least one SRAM, using existing on-chip SRAMs to construct PUFs is a very interesting solution, especially in IoT applications where resources are limited and constrained. An SRAM-based PUF relies on the start-up value (SUV) variability of SRAM cells as a source of entropy. The SUV of a single cell depends on two factors: on the one hand it depends on the intrinsic cell variability due to fabrication process; on the other hand, it depends on the specific operating conditions. The SUV repeatability of a given cell depends on environmental conditions such as temperature variations, aging, and noise. Error correction codes (ECC) can be used to decrease SUV variability to produce more reliable keys [7]. An ECC is calculated at first use and then it re-used as helper data to re-establish the reliable PUF output from the raw PUF output. However, regarding to the PUF output that have high error rates, some of these codes can be very complex increasing area overheads [8].

SUV repeatability can be also increased by classifying all SRAM cells between reliable and unreliable, and then using only the most reliable ones as SRAM-PUF cells. Using a cell subset formed by cells whose start-up value is more repeatable allows implementing a more reliable PUF. In [9,10], SRAMs cells are classified on the basis of their start-up reliability under different external condition variations such as supply voltage, ramp up time, and temperature. Exploring an appropriate range of these operational conditions in post-processing PUF reliability can be assessed [11]. With this technique, cells that show less ability to keep the same start-up value under different external conditions are recognized in such a way that they can be eliminated from the PUF implementation [9]. The main disadvantage of this technique is the potential high number of necessary tests that, in turn, increase test time and costs. In [8], an indirect selection method is presented, in which a test for each cell is performed, and those cells that pass the test are categorized as reliable and thus are candidates to be used as PUF generators. These kinds of tests are usually executed under single operational conditions, and one of their drawbacks is that they can detect some unreliable cells as reliable if conditions are not well chosen.

The paper's major contribution is to establish analytical reliability criteria for classifying the repeatability of SRAM cells for PUF implementation. The authors of the study propose the use of both introduced analytical metrics for estimating the number of SRAM cells useful for manufacturing PUF solutions and, during PUF design stages, to explore how cell design factors (such as threshold voltage and internal node start-up conditions) as well as environmental perturbations (such as temperature) could affect device performance. The percentage of reliable SRAM-PUF cells with more consistent SUV may be approximated using the proposed metrics, allowing the SRAM implementation design to be adapted and changed to fit the predicted proportion of PUF reliable cells to the PUF design requirements. Alternatively, the PUF design can be tweaked to match the SRAM's projected performance.

The rest of the paper is organized as follows. In the next section, some considerations are introduced about how SRAM instance is used to implement PUF applications. Section 3 defines separatrix and mismatch-based metrics used to estimate the percentage of suitable SRAM cells. Section 4 reports the results about the impact external and internal disturbances explored in this work. The reliable cells classification and the estimation of repeatable cells percentage are discussed in Section 5. Finally, Section 6 concludes the paper.

## 2. SRAM as PUF

A 6T SRAM cell consists of two access transistors, which are controlled by the word-lines (WL), and two cross-coupled inverters, which create a latch (Figure 1). If a memory is intended to be used to store data, and also as a PUF, it is necessary to be able to modify the power supply node ($V_{dd}$) to start up the cell. Having control over $V_{dd}$ provides a mechanism to start up the cell and thus challenge the circuits to provide a response that will be used by the PUF circuit. WL signal controls whether the memory cell is in hold state or if it can be accessed to perform read and write operations. Adding PUF functionality to a memory array requires minimal design modifications, especially in embedded memories that already include control over $V_{dd}$ to provide energy consumption savings solutions.

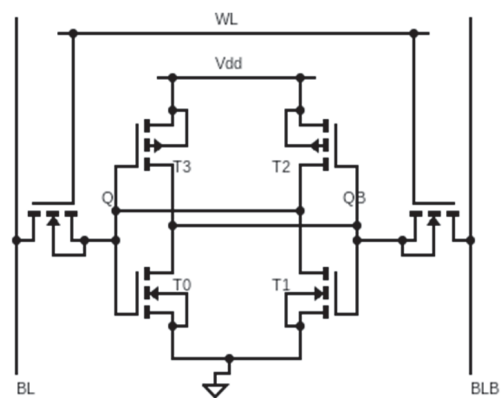

**Figure 1.** 6T SRAM cell schematic.

An SRAM cell has three equilibrium points: A and C, which are stable (Figure 2), and that correspond to the logic '1' and logic '0' states, respectively, and a third one (B), which is a meta-stable state point. It is located at the intersection between the two butterfly curves of each cell (Figure 2) [12]. Due to the mismatch between cell inverters, this meta-stable point location varies from cell to cell, as can be seen in Figure 2. Due to this, there are asymmetric cells that show a preference for point A or C during power-up.

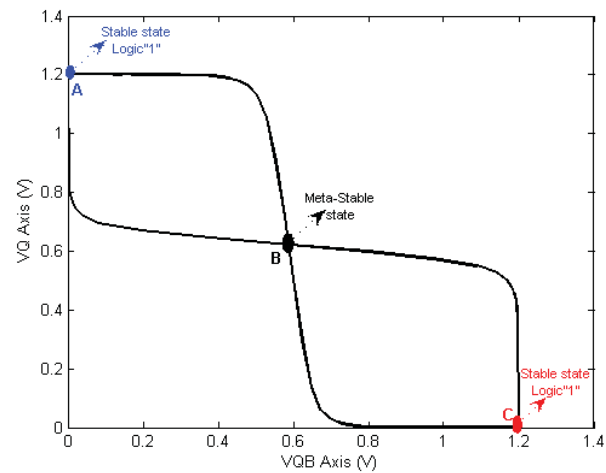

**Figure 2.** Butterfly plot of the SRAM cell. A and B are stable states, and C is metastable.

When $V_{dd}$ is raised (SRAM power-up), cells are kept in hold operation, i.e., access transistors are cut-off (WL = 0 V), and thus both internal cell nodes are isolated from the bit-lines. Provided that initially $V_{dd}$ = 0, both SRAM internal nodes are discharged (Q = QB = 0 V) and, as $V_{dd}$ is raised, Q and QB voltages will follow the power supply ramp, increasing its voltage. For low power supply voltages, the cell is near its meta-stable point, and thus it is not storing any value because it can still turn to any of the two possible final states ('1' or '0'). This situation continues until one of the internal nodes' voltage becomes slightly closer to one of the stable states than the other. When the power supply voltage reaches that specific value (known as flip-point), the final reached state will depend on several factors, such as the inverter transistor mismatch, the internal nodes voltage difference, and the amount of noise present.

This paper studies two different metrics aimed at identifying those SRAM cells that have one preferred stable state and thus will a have a tendency to start-up to one of the two stable states. These cells are expected to be more appropriate for reliable PUF implementations since their SUV are also expected to be more repeatable.

## 3. Metrics Definition
### 3.1. Mismatch-Based Metric

SRAM inverters imbalance caused by process variations mismatch is a major challenge that produces a major effect on noise margin in SRAM read and write operations.

Furthermore, a PUF is capable of producing a fingerprint that is unique to each integrated circuit. If the PUF is based on SRAMs, it is achieved relying on the intrinsic mismatch between the transistors that form the SRAM cell. One of the main sources of this mismatch is caused by random threshold voltage variability ($V_{th}$) [8].

A cell will start-up at Q = '1' if transistors T1 and T3 (Figure 1) are stronger than transistors T0 and T2, and vice versa for the SUV Q = '0' case. As a consequence, a high mismatch between cell inverters means a highly preferred stable state. Choosing the memory cells with the highest mismatch will lead to more repeatable and dependable SRAM-PUF fingerprint. This paper deals with the mismatch metric (MF) introduced in [13] as a predictor for SRAM memory cells SUV. The metric intended to be used at identifying those SRAM cells with the most repeatable start-up values, and thus better candidates for PUFs.

The MF measures the variation between SRAM cell inverters, as a function of the threshold voltage ($V_{th}$) of the four transistors that form the cross coupled inverters of the cell. The MF is defined as

$$MF = (|V_{th,3}| - |V_{th,2}|) - (V_{th,1} - V_{th,0}) \quad (1)$$

where $V_{th,n}$ represents the threshold voltage of each of the four transistors of the SRAM inverters. In Figure 1, each transistor is labeled with Tn, with n = 0, 1, 2, and 3. Notice that MF is defined as a subtraction of two terms: (i) the difference between T2 and T3, the two PMOS transistor $V_{th}$ ($\Delta P = |V_{th,3}| - |V_{th,2}|$), and (ii) the difference between T0 and T1, the two NMOS transistor $V_{th}$ ($\Delta N = V_{th,1} - V_{th,0}$). Both $\Delta P$ and $\Delta N$ quantify the mismatch between the same transistors type threshold voltage. Positive values of $\Delta P$ and negative values of $\Delta N$ are related to cells more prone to SUV Q = '0', while negative values of $\Delta P$ and positive values of $\Delta N$ correspond to cells with a tendency to SUV Q = '1'. As a consequence, the MF sign is correlated to the favored SRAM cell SUV value. If MF is positive, the left inverter has higher $V_{th}$ values than the right one (Figure 1); consequently, during a power-up, the right inverter will pull down the Q node earlier than QB, and the feedback mechanism will lead to a '0' SUV. Conversely, if MF is negative, the left inverter will pull down the QB node earlier than Q, and the cell will start-up to '1'.

One important factor to account for is that both transistor types (PMOS and NMOS) transistors do not play an equivalent role due to intrinsic differences, i.e., electron mobility and current drain capacity. For this reason, the previous definition of MF is modified to account for these differences between transistor types with a weight factor (WF). If Equation (1) is rewritten using this weight factor, as well as the $\Delta P$ and $\Delta N$ parameters, the following equation is obtained:

$$MF = WF \cdot \Delta P - (1 - WF) \cdot \Delta N \quad (2)$$

where WF is obtained by transient Monte Carlo simulations and applying an optimization process to adjust the MF parameter to the simulated SUV. A fitting process to optimize the number of cells whose SUV is equal to the sign of MF is performed, and after that, WF value of 0.765 is obtained. Figure 3 shows $\Delta N$ and $\Delta P$ values obtained with Monte Carlo simulations of an SRAM cells using a 65 nm commercial technology. Cells with SUV Q = '0' are red colored, while cells with SUV Q = '1' are blue. In addition, a line representing the MF = '0' values are also plotted and splits the figure into two zones. Figure 3 represents the MF values accounting for the WF optimization; in addition, the SUV values are represented by each dot color. Most cells located above the MF = '0' line (MF positive) start at '0' value, while most cells located below the MF = '0' line (MF negative) start at '1' value. Furthermore, there are a few cells that show an unpredicted behavior when MF is near 0 (close to the MF = '0' line). Despite this, all cells with high enough MF absolute values show an SUV that can be predicted by the MF parameter.

The MF value distribution is shown in Figure 4 in the form of a histogram. Cells with an MF factor located to the far left (blue bars) will start-up at '1', while cells with an MF located to the far right (green bars) will start at '0'. The rest of the cells, with an MF factor closer to zero show a more unpredictable SUV. As a consequence, the cells that are farther from MF = '0' show the most predictable SUV. We refer to the predictable SUV cells as reliable cells, and to the unpredictable SUV cells as unreliable PUF cells. In fact, from Figure 4, an MF threshold value can be derived to distinguish between reliable and unreliable cells. We define this threshold (see Figure 4) where the different histogram areas overlap. Using a 65 nm CMOS technology, the range overlap metric range is [−0.05 V, +0.05 V]. Out of this range, cells SUV behavior can be correctly predicted from the MF parameter.

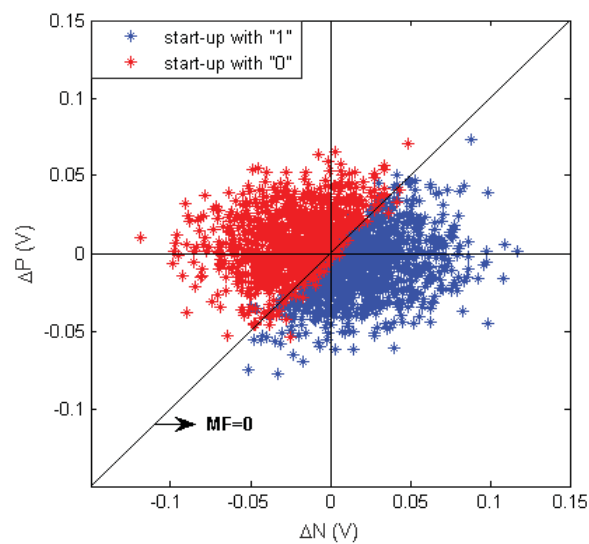

**Figure 3.** The distribution of threshold voltage fluctuations in NMOS and PMOS transistors.

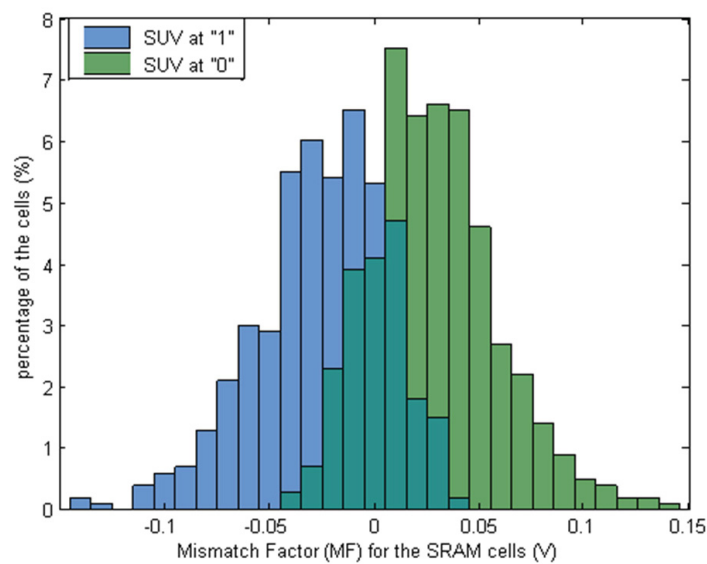

**Figure 4.** Mismatch factor histograms depicting the SUV in different colors. The unpredictable memory cells obtained with 65 nm CMOS 6T-SRAM can be observed by the overlap of histograms, which are all close to MF = '0'.

As a conclusion, SUV of cells with low MF absolute values are not accurately predicted with this metric. The fact that the SUVs of cells near MF = '0' are not correctly predicted is a consequence of the well-matched inverters that these cells have.

*3.2. Separatrix-Based Metric*

3.2.1. SRAM Separatrix Using State Space Representation

In [14], the SRAM dynamic noise margins (DNMs) were suggested. The write and read DNMs are defined and evaluated using the concepts of stability boundary and state-space separatrix, with the purpose of ensuring successful write and read operations. However, the authors indicate that the state-space separatrix might be used in hold operation because of the comparable dynamic features.

The state-space analysis is used to characterize the behavior of memory cells throughout their dynamic evolution [15], and this state-space analysis is represented by a second-order nonlinear time invariant system in [16]. A state-space analysis is a mathematical model that represents a physical system using inputs, outputs, and state variables that are associated with differential equations. Within the state-space, the system state can be represented as a vector. A phase-space is a space in which all of the system's potential states are shown, with each potential state referring to one of several locations in the phase-space. The trajectory is the non-forced progression of each potential state to the equilibrium point (black dotted lines in Figure 5).

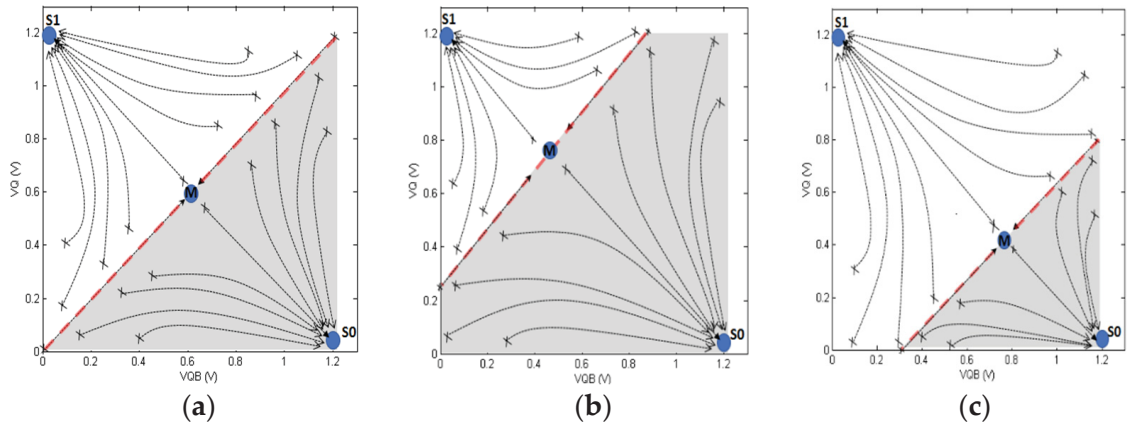

**Figure 5.** The phase-space of SRAM memory development at startup: (**a**) for an ideal symmetrical cell, (**b**) for an asymmetrical cell with a propensity towards logic '0', and (**c**) for an asymmetrical cell with a tendency towards logic '1'.

As previously mentioned, the SRAM cell has three equilibrium points, two of which are stable and represent logic '0' (VQ = 0 V, VQB = $V_{dd}$) and logic '1' (VQ = $V_{dd}$ V, VQB = 0 V) states, S0 and S1, respectively, in Figure 5. The other is a meta-stable condition, which is represented by point M (blue circle between S1 and S0) in Figure 5. Each stable point in the state-space has its own area of attraction, and the border between these areas is known as the SRAM separatrix (red dashed lines in Figure 5). If a memory cell is initiated from any of the beginning node conditions (VQ$_o$, VQB$_o$) in the region of attraction, its state trajectory will trend to one of the stable-state points as time passes. In a perfect cell, on the other hand, if the cell is started up from any beginning condition located on the separatrix line, the state trajectory will travel to the meta-stable state. However, the influence of process variation and mismatch will drive the state-trajectory of a genuine memory cell towards one of the stable states. The SRAM separatrix is the barrier between these locations in state space (red dashed lines in Figure 5).

Therefore, if the cell is started up from any initial condition positioned on the separatrix line, the state trajectory will travel to the meta-stable state, as shown in Figure 5a. The influence of process variation and mismatch, on the other hand, will drive the state trajectory of a non-ideal memory cell towards one of the stable states.

The presented phase state in Figure 5b corresponds to a mismatched SRAM cell, which has a larger region of attraction to the S0 state (shadowed region) than to the S1 state (logic '1'). As a result, it will prefer to begin at logic '0'. Figure 5c, on the other hand, depicts a cell with a larger region of attraction to the S1 state (unshadow area) and hence prefers to begin at logic '1'. The strength of the memory cell's intrinsic mismatch is determined by the size difference between the zones of attraction in each scenario. To put in other words, the amount of intrinsic cell mismatch is defined as the distance between the separatrix line and the ideal location at VQ = VQB line.

A stability test for PUF applications has been developed in [17], which focuses on the dynamic evaluation of memory cell stability. On the basis of previous results, we use in this work a two-step test approach, as follows:

- The memory is initiated with the cells in the stable state S1 as their initial condition: setting node QB$_o$ = 0 V and node Q$_o$ = $V_{skew}$.

- The memory is set up again, but this time with cell starting conditions skewed towards the stable state S0: setting node $QB_o = V_{skew}$ and node $Q_o = 0$ V.

When a cell's SUV in the first tests is S1 and its SUV in the second test is S0, the cell is said to be strongly matched. By contrast, if a cell's SUV is the same in both tests, it is termed a mismatched cell, which means it will be extremely stable in PUF applications. Even while this stability test can categorize highly symmetrical cells, if the selected value for $V_{skew}$ is low, the method's primary shortcoming is picking the initial condition value ($V_{skew}$). Some of the unreliable cells may be undetected by the test due to an extremely low $V_{skew}$ value, while a high value of $V_{skew}$ can lead to oversizing the needed memory for PUF implementation, as more reliable cells will be deleted and designated as unreliable cells.

3.2.2. Separatrix Intersection Distance (SID) Metric

In this section, we offer a novel metric-based technique to categorize SRAM-PUF cells on the basis of their dependability. Each cell in the proposed memory is assigned a value on the basis of the position of the separatrix line. In this case, if the separatrix line for a mismatched cell is above the ideal line (symmetrical cell); as shown in Figure 5b, the cell will have a tendency to power up at logic '0'. If the separatrix is below the ideal line, on the other hand, the cell will prefer to power on at logic '1', as shown in Figure 5c. Therefore, the suggested metric seeks to determine how distant the separatrix is from the symmetric position. The further a cell's separatrix is from its optimum position, the wider the region of attraction towards one of the cell's stable states will be, and therefore the more reliable it will be. Specifically, the separatrix of a cell with a predisposition to logic '0' state will only intersect the Q-axis (see Figure 6a), whereas the separatrix of a cell with a predisposition to logic '1' state will only cross the QB-axis (see Figure 6b). Separatrix intersection distance (SID) is a new metric that will be defined as the distance between those intersection points and the intersection point of the ideal separatrix at ($VQ_o = VQB_o = 0$ V), as illustrated in Figure 6.

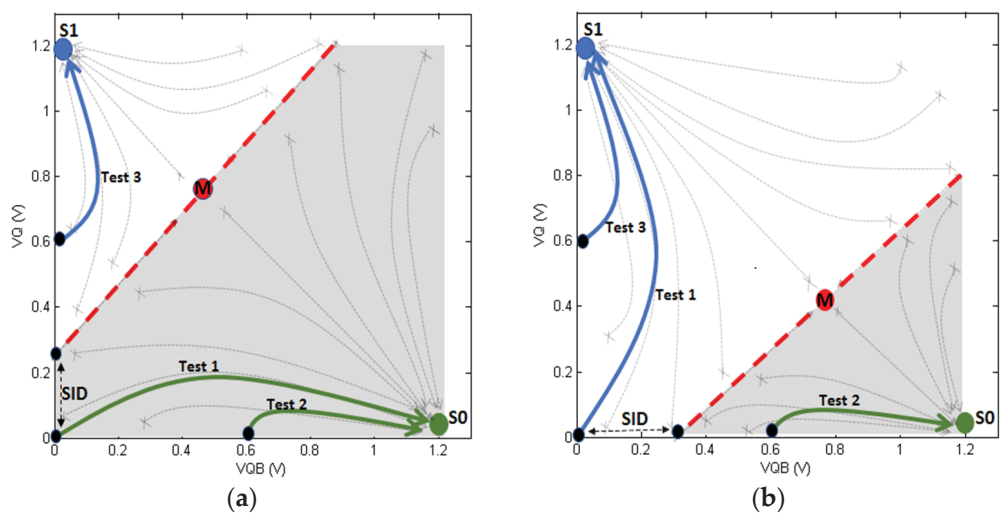

**Figure 6.** The suggested tests and SID metric are provided in the phase space of memory cell evolution: (**a**) for an asymmetrical cell with a propensity towards logic '0', and (**b**) for an asymmetrical cell with a tendency toward logic '1'.

The new SID metric is evaluated using transient Monte Carlo analysis. The approach adopted may be separated into two sections: the first one is devoted to determining the direction of the SUV tendency (either to logic '0' or logic '1') by detecting where the separatrix crosses the phase state axis (either at Q-axis or QB-axis), while the second part establishes the intersection value. The following is a summary of the technique used in this approach:

- Three tests for each cell in the proposed memory are implemented to determine which axis the separatrix would intersect:

- Test 1: VQB = 0 V and VQ = 0 V are the initial conditions for the cell, as shown in Figure 6.
- Test 2: the cell is powered on with starting conditions highly skewed towards the stable state S0 (logic '0'), with VQB = $V_{dd}/2$ V and VQ = 0 V (see Figure 6). The initial $VQB_o$ value is set to identify the trend toward S0, even in highly symmetrical cells; a low initial condition value does not allow for the detection of low mismatched cells [18].
- Test 3: the cell is powered on with initial condition highly tilted towards the stable state S1 (logic '1'), where VQB = 0 V and VQ = $V_{dd}/2$ V (see Figure 6). Again, the initial $VQ_o$ value is set to detect the trend towards S1.

Following the performance of those tests using the test memory instance, this work discovered that if a cell's final SUV in Test 1 is identical to its final SUV in Test 2, the cell will gravitate toward S0, and its separatrix will intersect the Q-axis, as shown in Figure 6a. In addition, if the final SUV of a cell in Test 1 is comparable to the SUV in Test 3, the cell will start up at S1 and the separatrix cross will be on the QB-axis, as presented in Figure 6b. Afterwards, the value of intersection on the detected axis is obtained. This value, SID, in both Figure 6a,b, represents when a cell starts changing its evolution towards one stable state (S0 or S1). To find the separatrix crossing value, we implemented a search algorithm in Ocean platform. This algorithm explores from the initial condition (starting at point VQ = VQB = 0 V) to the detected node (Q-node or QB-node that is decided in the previous step) in steps of 5 mV. Until the start-up changes, the algorithm will keep altering this initial state while checking the SUV. SID stands for the value of the starting condition for the identified node when the startup behavior changes.

Positive SID values are assigned to the cells that have a tendency to logic state '0' (these are the cells that have intersection with the Q-axis, as shown in Figure 6a), while the sign is negative for the cells with a tendency to logic state '1' (intersection with the QB axis, Figure 6a). The histogram of the SID values obtained for each cell is shown in Figure 7. The most reliable cells are located at the extremes of the histogram, since they have the largest SID values, and therefore their separatrix are shown far from the ideal position and thus have a larger area of attraction towards one of the stable states of the cell.

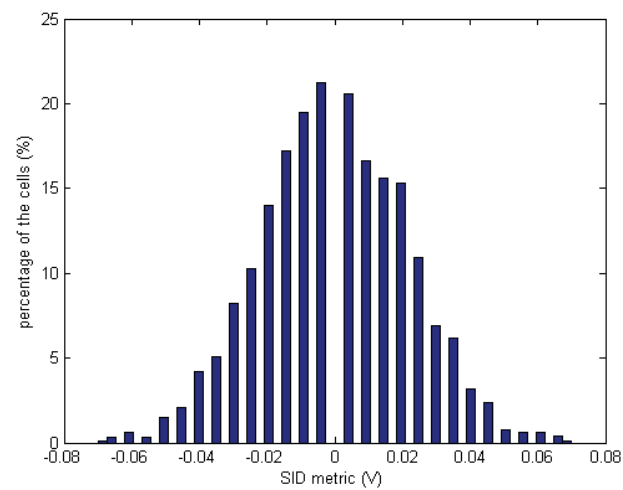

**Figure 7.** The histogram of the SID metric.

*3.3. Relationship between the Proposed Metrics*

The proposed metrics attempt to clearly identify the cells that exhibit the best operating conditions for use in PUF applications. Thus, memory cells that present a high absolute SID value should show a high mismatch metric, while memory cells with balanced inverters should offer low absolute values in both metrics, denoting low repeatability of the SUV. Figure 8 presents the relationship between the proposed SID metric and the inherent mismatch represented by MF metric. Each star in the figure represents a different memory cell. The black color represents cells that have an initial SUV towards the logic state '0', and the red color represents cells that have an initial SUV towards the logic state '1'. There is a high correlation between both metrics, which means that the further the distance of the separatrix from the ideal, the higher the mismatch the cell will present. This correlation has a linear coefficient equal to 0.965, which shows the agreement of both metrics in estimating the classification of the memory cells for PUF implementation.

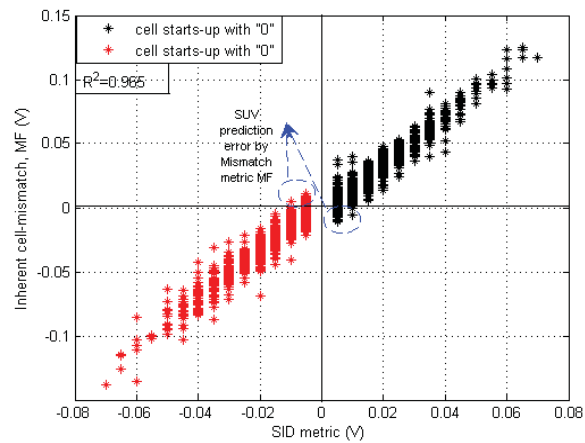

**Figure 8.** The relationship between the separatrix metric (SID) and mismatch metric, in which the coloring is based on the original SUV set as a reference.

## 4. The Impact of External and Internal Disturbances on SRAM-PUF

In terms of reliability, PUFs should deliver the best possible outcomes. In the case of SRAM-PUF cells, this may be accomplished by employing cells with highly repeatable start-up value, even with the presence of external disturbances (temperature changes for example) or internal noise (thermal noise).

The statistical behavior of a PUF may be assessed in two ways: the first is by testing numerous instances of the circuit and obtaining its reliability; the second is by modeling the SRAM cell noise and simulating its statistical behavior. To show the performance of the suggested metrics, we used 65 nm commercial CMOS technology in this research.

### 4.1. Noise Modeling in SRAM-PUF

The internal thermal noise that exists in SRAM cells can significantly modify the SUV. Emulation of the actual startup behavior of an SRAM thermal noise in the memory cells can be modeled by inserting random transient voltage sources between the cross-coupled inverter storage nodes of the cells following a configuration scheme similar to the one described in [19]. The complete schematic of the memory cell model with the noise voltage sources is shown in Figure 9. The magnitude of the thermal noise at any node should be characterized in terms of a normal distribution with zero mean, while the standard deviation of this noise is mainly based on the node capacitance (*C*) and temperature (*T*) from the following equation [18,19]:

$$\sigma_{noise} = \sqrt{\frac{KB\,T}{C}} \quad (3)$$

where *KB* is Boltzmann constant. In [18], the standard deviation of internal thermal noise for each cell node in the memory was set to 4.5 mV in 90 nm CMOS technology to produce enough SUVs variability level. Accordingly, we set $\sigma_{noise}$ to 8.5 mV for each cell node using 65 nm CMOS technology.

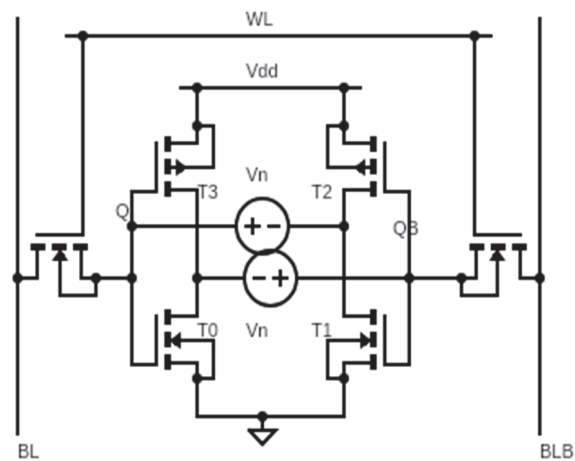

**Figure 9.** SRAM cell including two random noise voltage sources.

When noise was present, some cells switched SUVs every time they started up, whilst others have a more consistent SUV. The SRAM cells were successively powered up 200 times using Monte Carlo simulations to observe the repeatability of SUVs. Prior to power-up of the memory cells, it is necessary to ensure that both Q and QB cell nodes are completely discharged. Each memory cell's statistical SUV behavior is calculated and expressed as the probability of starting up in one of two states: logic '0' or logic '1'. Figure 10 shows a graphic representation of the chance of SUVs as a result of using this technique, where the color variations denote the preferred SUV probability.

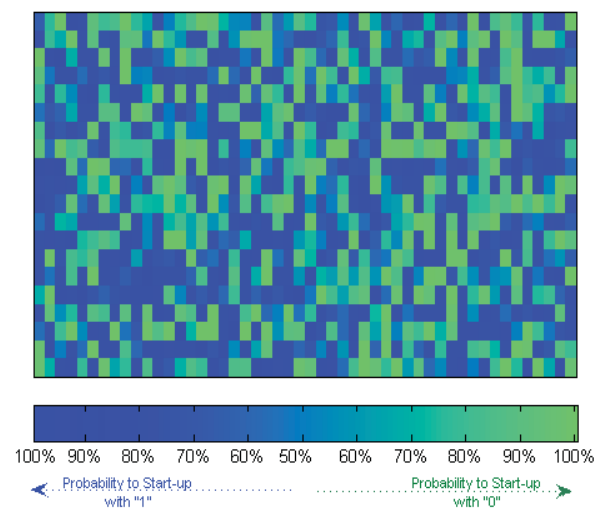

**Figure 10.** Visualization of the probability of each memory cell to start up to a preferred SUV.

Memory cells with greener color have a higher probability to starting up in logic '0' direction, whereas memory cells with a deeper blue color have a higher probability to start-up in logic '1' state.

On the other hand, memory cells that have about 50% SUV probability are represented by colors located at the center of the legend in Figure 10. The start-up behavior of these cells is greatly affected by internal noise and may therefore be considered as memory cells with random SUV. The repeatability of the SUVs is associated with the immunity of SRAM cells against internal noise and reflects the strength of the inherent mismatch of the memory cell; thus, a highly mismatched cell is able to repeat its SUV on every power-up. To corroborate this, this study analyzed the relationship between the probability of repeating SUV (probability of starting at logic '0' or '1') and the proposed metric. The results show that well-matched cells with low absolute MF values had very low repeatability, and their probability of starting at '0' or '1' was around 50% (random SUV), while strongly mismatched cells could have a higher repeatability, even when considering the existence of thermal noise. This result can be seen in Figure 11, where the correlation between the SUV probability and the proposed metrics is calculated for each memory cell. Each cell is represented by one star, where the color is based on the SUV probability—the greener the memory cell, the higher the probability of starting towards logic '0' (higher P('0')), and the bluer the memory cell, the higher P('1') it will have; P('1') = 1 − P('0'). Therefore, both sets of probabilities are complementary. However, the SUV probability is closely related to the value of the proposed metrics. We notice in Figure 11 that memory cells with a higher P('1') had the most negative MF and SID values, while memory cells with a higher P('0') had the most positive MF and SID values. This correlation between SUV probability and the sign of the proposed metrics is in line with previous sections assumptions where a higher absolute value of the metrics indicates more repeatable SUV and thus more reliable cells.

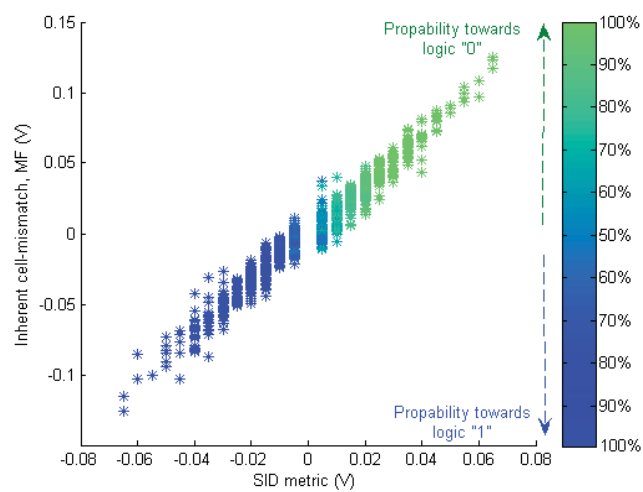

**Figure 11.** The relationship between the probability of the cell to starting up to a preferred state and the proposed metrics.

This demonstrates the ability of the metrics to identify most memory cells that have repeatable SUVs. In addition, cells that have a very low SUV probability (around 50%) may also be identified and used for random number generation applications since the probability distribution of repeating start values is physically random.

*4.2. Impact of External Temperature Variations*

Temperature is another well-known external perturbation that affects the SUVs repeatability. The impact of this parameter on the reliability of SRAM-PUFs has been extensively explored in the literature [20,21]. According to these studies, temperature fluctuations affect the start-up behavior of memory cells, and the selected PUF cells should exhibit stable SUVs under such fluctuations.

In this study, the effect of temperature to identify stable and unstable cells is investigated, with the main objective of demonstrating the ability of our suggested metric to categorize the suitable cells. Therefore, unstable cells will be characterized as memory cells that alter their SUVs at any temperature in the simulated range, while stable cells will be defined as memory cells that tolerate temperature fluctuations without changing their SUVs.

The temperature range explored starts from −40 °C and reaches up to 120 °C, in steps of 20 °C. Figure 12 compares the percentage of memory cells that change their SUVs at any temperature in the range, with the SUVs achieved with the typical corner technology and nominal temperature (27 °C).

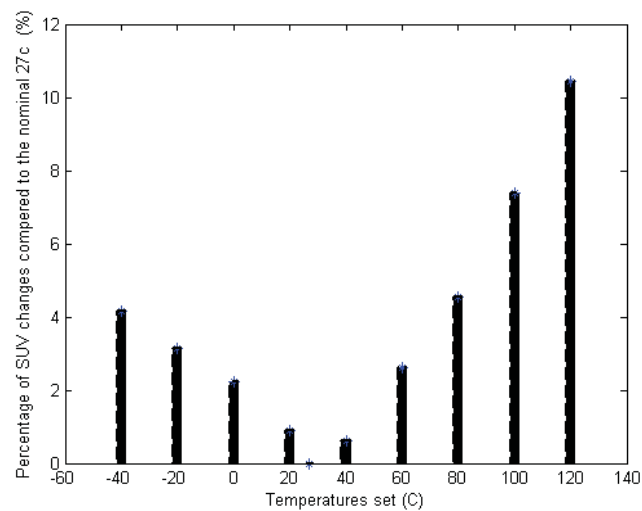

**Figure 12.** Percentage of cells changing SUV due to temperature variations from nominal 27 °C.

This paper aims to examine how non-specific SRAM design may be used to generate high reliability cells for PUF applications. As a result, the memory can experiment a rise in temperature produced by long runtime periods, and cells that alter their SUVs as a result of this increment in temperature should be flagged as unstable cells and masked out from PUF response. Useful cells keep the integrity of their SUVs, and they will be regarded as stable. Similar to the previous section, Figure 13 highlights the stable cells (gray colored) in the relationships between novel metrics.

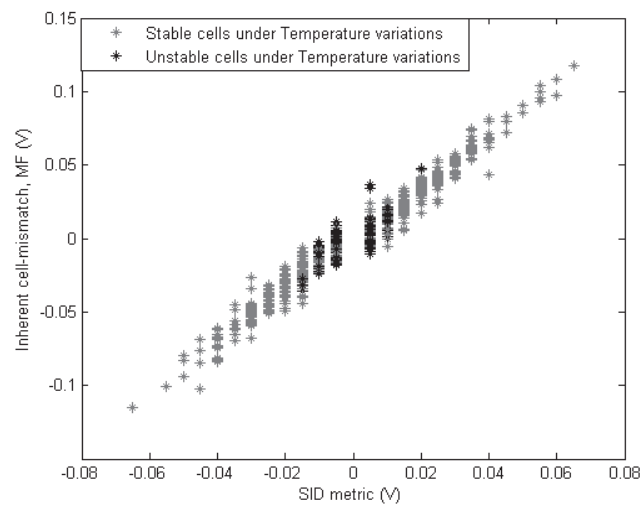

**Figure 13.** The relationship between the proposed metrics demonstrating stable and unstable cells under temperature changes.

Although temperature affects roughly 11% of memory cells, the majority of these cells are clustered at the lowest absolute values of the suggested metrics. As shown in Figure 13, the SID and MF metrics are capable of categorizing the SUV stability against temperature swings. The cells with the greatest absolute values of the metrics have more consistent SUV under temperature fluctuations.

## 5. Identification Reliable PUF Cells Using the Proposed Metrics

According to the previous results, a reliable cell is defined as the one that has highest probability to repeat the SUV as long as it has a stable SUV under temperature fluctuations. The reliable cells are particularly suitable for PUF implementation since their SUVs are more consistent and stable in presence of internal noise and external temperature variations.

Figure 14 depicts the MF and SID metrics' histogram distributions. On both metric histograms, the reliable and unreliable cells are emphasized. Figure 14a shows the classification obtained using mismatch metric (MF), whereas Figure 14b represents the categorization of memory cells using SID metric. The proportion of reliable cells is shown by red bars, while the yellow bars show the percentage of the unreliable cells. The reliable cells have the highest absolute values using both metrics. Unreliable cells are also identified by their low metrics magnitude. These results indicate that these metrics are effective at classifying PUF cells.

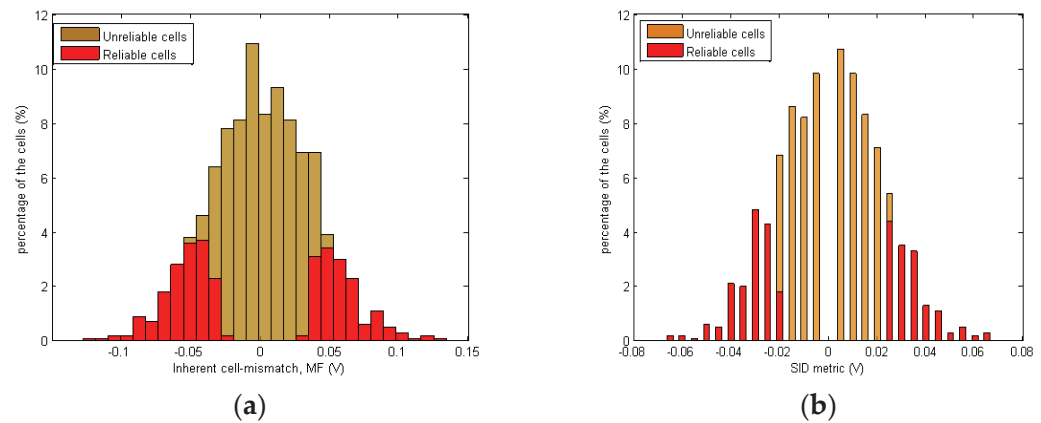

**Figure 14.** The histograms for the proposed metrics indicate a consistent distribution: (**a**) for MF metric, (**b**) for SID metric.

On the basis of the concept that greater absolute values of the metrics may greatly increase SRAM-PUF reliability, the suggested approach can easily discover the most reliable PUF cells. In this study, a 256 PUF responses were selected from the whole intended RAM. The memory cells with the largest metric magnitudes were chosen for the 256 PUF-bits. The selection results for proposed metrics are summarized in the Table 1, where random cell selection was introduced to support the effectiveness of the metrics.

**Table 1.** Selection results.

| Identification Methodology | Stable Temperature Cells | Cells with Highest Probability to Repeat the Same SUV | Reliable Cells |
| --- | --- | --- | --- |
| MF (256 bit) | 100% (256 bits) | 95.7% (245 bits) | 95.7% (245 bits) |
| SID (256 bits) | 100% (256 bits) | 98.8% (253 bits) | 98.8% (253 bits) |
| Random cells (256 bits) | 76.6% (196 bits) | 33.2% (85 bits) | 30.9% (79 bits) |

Each column of this table shows the number and percentage of cells selected as stable under temperature variations, the cells with 100% probability to repeat the SUV considering internal noise, and the reliable cells (memory cells that meet the highest probability to repeat the SUV and are stable SUV under temperature fluctuations).

The presented measures reveal that the selected PUF cells had a good categorization with a high percentage of reliable cells. The SID metric obtained the best identification results; however, the MF measure came close. It is also worth noting that 86.3% (221 bits) of the chosen cells were shared by both metrics. When we compared the acquired results to random cell selection, we were able to see how the metric technique enhanced SRAM-PUF dependability when the cells with the best metric were chosen.

As previously stated, the reliable cells were more concentrated at the greatest metric values, and their concentration reduced as the metric value fell. Starting with the greatest metric absolute values, the PUF cells can be selected for highest reliable implementations. As the challenge-response length of PUF determines the number of SRAM cells required, if a low number of SRAM cells are required, a greater number of reliable cells will be used to construct the response, increasing the PUF dependability. Figure 15 shows the percentages of reliable cells identified by each metric methodology for various bit lengths (128, 256, and 512 bits) for constructing the SRAM-PUF response. When the response length is shortened, the proportion of detected reliable cells clearly increases.

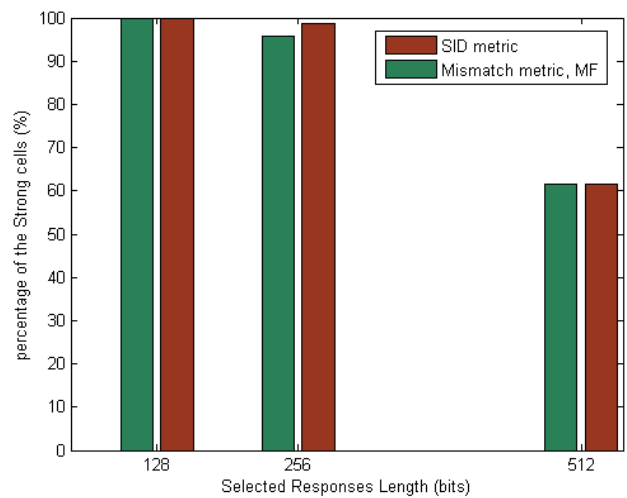

**Figure 15.** Percentage of reliable cells identified by both metrics when changing different response lengths.

## 6. Conclusions

New reliability metrics for SRAM-PUF are introduced in this study, allowing for the estimation of the percentage of strong and weak cells that will be available in an SRAM memory. One of the advantages of these metrics is that they can be obtained at early stages of SRAM design. Results were validated comparing them to the start-up values (SUV) obtained by Monte Carlo simulation. To perform this analysis, we generated a set of cells affected by process variability—each cell is then analyzed to determine its probability to start up at a given logic state in the presence of realistic temperature fluctuations and random noise. We observed that both proposed metrics, MF and SID, are appropriate for the identification of those cells that can be used in a PUF module. The low cost in complexity and computation time of the proposed metrics (relative to SUV) makes them suitable for being applied in the SRAM design phases to obtain a good estimation of how many cells may be suitable for using in a PUF in a given SRAM design.

**Author Contributions:** Conceptualization, A.A. and S.A.B.; methodology, A.A. and G.T.; validation, A.A., G.T. and B.A.; investigation, A.A.; resources, B.A.; writing—original draft preparation, A.A.; writing—review and editing, B.A., G.T. and S.A.B.; supervision, G.T., S.A.B. and B.A.; funding acquisition, B.A. All authors have read and agreed to the published version of the manuscript.

**Funding:** This research was funded by University of the Balearic Islands and IBETEC under the project IOTIB/3739.

**Conflicts of Interest:** The authors declare no conflict of interest.